# Effects of H⁻ low beam irradiation and high field pulsing tests in different metals

C. Serafim,[1,2,*] S. Calatroni,[1,†] F. Djurabekova,[2] R. Peacock,[1] V. Bjelland,[1,3]
A. T. Perez-Fontenla,[1] W. Wuensch,[1] A. Grudiev,[1] S. Sgobba,[1]
A. Lombardi,[1] and E. Sargsyan[1]

[1]*CERN, European Organization for Nuclear Research, 1211 Geneva, Switzerland*
[2]*Helsinki Institute of Physics and Department of Physics, P.O. Box 43, University of Helsinki,
FI-00014, Finland*
[3]*Department of Physics, NTNU, 9491 Trondheim, Norway*

This work studies the suitability of a set of different materials for manufacturing more efficient and durable radio-frequency quadrupole (RFQ) structures compared to that currently used in many linear particle accelerators, traditionally made out of copper. RFQs are susceptible to vacuum breakdowns caused by the exposure to high electric fields, resulting in surface degradation. Additionally, a further limitation of present-day copper RFQs is surface blistering under hydrogen ion beam exposure, due to beam halo losses. Irradiation is associated with a further reduction of the breakdown field strength of the metal surface thereby affecting the overall efficiency of the RFQ. The investigated materials, Cu-OFE, CuCr1Zr, CuBe2, Ti6Al4V, SS316LN, Nb, and Ta, were submitted to low-energy (45 keV) H⁻ irradiation and tested in a direct-current (dc) system with pulsed high voltage. For comparison, the maximum surface electric field was measured for both irradiated and pristine (nonirradiated) surfaces of the same material. The effects of irradiation on the surface of the materials, before and after being submitted to high electric fields, were studied with the help of microscopic imaging and chemical analysis. Blistering caused by H⁻ irradiation has been observed in all copper and copper alloy materials. Despite reductions in breakdown field strength postirradiation, no indications were found that the blisters on the surface have a direct cause in triggering breakdowns during high electric field tests. SS316LN, Ti6Al4V, CuBe2, and CuCr1Zr showed maximum surface electric fields higher than copper, making these promising candidate materials for future RFQs manufacturing. This paper focuses on the results with CuCr1Zr, CuBe2, SS316LN, and Ta, complementing and expanding previous work exploring Cu-OFE, Nb, and Ti6Al4V.

DOI: 10.1103/PhysRevAccelBeams.28.013101

## I. INTRODUCTION

The radio-frequency quadrupole (RFQ) is an important section within low-energy linear ion accelerators. Due to the radio-frequency (rf) component of the RFQ, which allows beam focusing, bunching and accelerating, it is possible to achieve high ion beam intensities in a very compact structure. An RFQ has a geometry of four electrodes, called "vanes," which form the rf cavity; detailed information about the vane fabrication, machining, and brazing can be found in [1].

*Contact author: catarina.serafim@cern.ch
†Contact author: sergio.calatroni@cern.ch

During the RFQ operation, the possible occurrence of rf electric breakdowns is a factor that can diminish the performance of the structure, leading to beam losses, and limit the lifetime of the RFQ. The increase of breakdowns in the RFQ cavities is correlated with their maximum surface electric field [2]—in the case of the RFQ at CERN LINAC4, the surface field in the structure is around 35 MV/m [3]. This becomes a limitation in the current RFQs and in the future developments for the reduction in size of these complex structures, as smaller dimensions mean a higher vane voltage and therefore higher surface electric fields that can give rise to higher breakdown rates. In the literature, there are no indications that breakdowns are associated with one single material property, and as reported in [4], there is not yet a fully comprehensive explanation of how they can be predicted.

A further important aspect of RFQs is the problems linked to their exposure to accelerator beams. Typically, when the structures are exposed to low-energy hydrogen ions irradiation (in case of proton or H⁻ beams), hydrogen ions can accumulate beneath the material's surface and recombine into molecular hydrogen. As hydrogen accumulates, bubbles are formed and the increase of their pressure inside the material may lead, depending on the material, to the formation of blisters on its surface. This phenomenon is known as blistering and is strongly dependent on the energy and flux of the beam, temperature, and material properties [5]. Blistering can affect the mechanical properties of the materials leading to surface degradation and failure, which will consequently affect the operation efficiency and longevity of the accelerator components. To try to minimize the surface damage caused by blisters, material properties such as permeability, diffusivity, and solubility were taken into consideration when choosing the materials to be tested in this study.

This work is based on the motivation to tackle these two main problems, which have been the primary concern in RFQ's material degradation within the first years of operation, by studying different materials and potentially finding better alternatives for future RFQ developments in terms of high electric field strength and beam loss resilience. Until now, no studies have been conducted to understand whether the blistering effects on the RFQ surface could be potentially linked to the appearance of vacuum breakdowns. In this study, this question is addressed by combining two different set of tests, with the first being high voltage dc pulsing, in a dedicated setup which enables us to replicate RFQ surface electric fields and study the breakdowns triggered from it in a controlled environment. And second, $H^-$ low- energy irradiation in order to see if the irradiated materials perform differently from the nonirradiated materials when submitted to the high-voltage test. The mentioned tests were made on electrodes from various materials (Cu-OFE, CuCr1Zr, CuBe2, Ti6Al4V, SS316LN, Nb, and Ta). For each material, two pairs of cathode and anode were machined with requirements aimed at achieving a face surface flatness of 0.001 mm and maintaining a surface finish below 0.025 μm of roughness. In essence, the copper alloys have been selected for their stronger mechanical properties compared to copper while retaining basically a similar behavior with respect to hydrogen. The other materials have been selected for their very high diffusivity, with Ti6Al4V having high mechanical strength. Stainless steel sits across the two categories: the interest for testing stemming from the fact of having displayed the highest field-handling capabilities in [6] and being a common construction material in particle accelerator technology. Detailed information about the materials can be found in Sec. II A. Details on the experimental setups and procedures are reported in Sec. II B. Complete experimental results are reported in Sec. III, noting that some preliminary results were reported in [7] and are presented here again for the sake of giving a complete picture of this study. Finally, it should be emphasized that it is the first time that such testing is performed with the chosen materials.

## II. MATERIALS AND METHODOLOGY

### A. Materials

In total, seven materials were chosen to conduct the research: pure oxygen-free copper (Cu-OFE), UNS[1] type C10100; copper chromium zirconium (CuCr1Zr), UNS type C18150, with approximately 0.1% of zirconium, 1% of chromium, and with the rest being copper; copper beryllium (CuBe2), UNS type C17200 with 98 wt% copper, and 2% beryllium; stainless austenitic chromium-nickel-molybdenum steel with nitrogen addition (AISI 316LN), UNS type S31653; premium-grade titanium alloy (Ti6Al4V, grade F-5, 6 wt% aluminum-4 wt% vanadium, UNS type R56400), forged in an $\alpha$–$\beta$ range with a final $\alpha$-microstructure; high purity (99.9%) niobium (Nb), UNS type RRR300% and 99.9% tantalum (Ta), UNS type UNS R05200. The chemical composition of each material alloy can be found summarized in Table I.

#### 1. Physical and mechanical properties

The choice of material, in terms of RFQ design, is crucial as it will significantly influence the efficiency, stability, and lifetime of the machine. Several parameters were taken into account when choosing the materials. Electric conductivity is important, as higher conductivity minimizes power losses, which means that more rf power is used for accelerating the beam, instead of dissipating as heat. A material with higher thermal conductivity is simultaneously as important, as the structure integrity of the vanes depends partially on the capability of the material to dissipate the heat generated by the rf power and beam interactions. During operation, RFQ vanes also experience mechanical stress caused by thermal expansions of the material and electromagnetic forces. For this reason, having a material with high mechanical strength in order to prevent deformations can help maintain the precise geometry of RFQ vanes. Material properties for the thermal and electric conductivity and mechanical strength are listed in Table II, according to the cited literature.

If one would focus only on electric and thermal conductivities, Cu-OFE or the Cu alloys would typically be the materials of choice due to those superior values. In comparison with the high-purity oxygen-free copper, the thermal conductivity of CuCr1Zr stays a bit below. However, this alloy is commonly used instead of copper for applications that require a higher mechanical strength, hardness, and thermal stability [8]. Also, it presents high electric conductivity, the second biggest among all chosen materials. The addition of Cr can improve the high mechanical strength of the alloy, and according to [25], it consequently promotes the reduction of dislocations in the material. CuBe2, showing lower values in the copper family for thermal and electric conductivity, represents a good compromise by having very good mechanical properties and hardness, with the highest mechanical strength among all materials. It s thermal expansion closely matches the steels, including the stainless grades [26,27].

---

[1]UNS stands for Unified Numbering System, typically used worldwide as an alloy designation system.

TABLE I. Chemical composition of the selected material alloys (wt%).

| | | | | | | | | | | |
|---|---|---|---|---|---|---|---|---|---|---|
| CuCr1Zr | Cu<br>Rest | Cr<br>0.8 | Zr<br>0.09 | Si<br><0.01 | Fe<br><0.01 | | | | | |
| CuBe2 | Cu<br>Rest | Be<br>2.0 | Co<br><0.30 | Ni<br><0.30 | Fe<br><0.20 | | | | | |
| SS316LN | Fe<br>Rest | Cr<br>17.24 | Ni<br>13.31 | Mo<br>2.52 | Mn<br>1,55 | Si<br>0,41 | N<br>0.16 | C<br>0.025 | P<br>0.023 | S<br>0.004 | Co<br>0.048 |
| Ti6Al4V | Ti<br>Rest | Al<br>6.06 | V<br>4.02 | Fe<br>0.11 | $O_2$<br>0.11 | C<br>0.0149 | $N_2$<br>0.003 | $H_2$<br>0.0046 | | | |

Ti6Al4V combines low density, high strength, and excellent corrosion resistance. This alloy has one of the highest mechanical strength values and lower thermal expansion, which is also important to assure that possible structural vanes can remain stable under varying thermal stress, and superior corrosion resistance in many harsh environments. Also, SS316LN has similar characteristics. The high achievable breakdown field of stainless steel material, reported in [6], represented a good indicator that this material could be a good candidate for the study, even with lower thermal conductivity when compared with copper and copper alloys. The presence of chromium in steels has been associated with a smaller dislocation mobility [6]. Pure Cr particles are known to provide high mechanical strength by dislocation motion inhibition to the material [25,28]. Both of these materials, Ti6Al4V and SS316LN, also offer good machinability but require careful handling when compared with copper and copper alloys, which are easier to machine into complex shapes as required for the RFQ vanes.

In terms of mechanical properties, Ta and Nb fall short when compared with all the other materials and additionally have difficult machinability due to their ductility. On the other hand, both Ta and Nb represent the strongest materials in terms of hydrogen embrittlement and therefore have been chosen to better explore their behavior in environments involving hydrogen ion beams. This brings us to one of the most relevant parameters that were also taken into account when choosing the materials, which is the ability of hydrogen diffusion in the material lattice.

### 2. Hydrogen diffusivity

It is well known that hydrogen diffusion in metals is highly associated with a decrease in mechanical and structural properties, due to the occurrence of phenomena such as hydrogen embrittlement or blistering.

Interstitial diffusion happens when the hydrogen atoms diffuse into the bulk material and occupy the interstitial sites in the metal lattice. The diffusivity, i.e., the hydrogen diffusing flux along the metallic material, follows Fick's law: $J = -D\nabla c$, where $J$ is the flux of hydrogen atoms, and, $D$ and $\nabla c$ are the diffusion coefficient and concentration gradient, respectively, of hydrogen in the metal. $D$ is a key characteristic of the migration of hydrogen in a metal, which is a quantitative characteristic of the rate of diffusion [29]. This thermodynamic parameter, follows Arrhenius dependence on temperature, according with Eq. (1), where $D_0$ is a constant, $E_d$ is the activation energy of diffusion (J/mol), $R$ is the gas constant (8.314 Jmol⁻¹ K⁻¹), and $T$ is the absolute temperature (K) [29].

$$D = D_0 \exp(-E_d/RT). \quad (1)$$

Another key factor of the process is the solubility factor in the bulk material, describing the maximum amount of hydrogen that can be dissolved in a metal lattice. The quantification of solubility is usually expressed as the Sievert's constant which is defined by the equation: $K_s = c/\sqrt{p}$, where $c$ is atomic gas concentration in the bulk volume and $p$ is the partial pressure of hydrogen gas [30]. Like diffusivity, also Sieverts's constant shows an Arrhenius dependence on the temperature [Eq. (2)]:

$$K_s = S_0 \exp(-E_s/RT). \quad (2)$$

At last, permeability ($\phi$) quantifies the ability of hydrogen to enter and move within the material, and it is given by the linear relation of the two concepts of diffusivity and solubility:

TABLE II. Compiled properties of the selected materials at room temperature (300 K) [8–24].

| Material | Electrical conductivity (MS/m) | Thermal conductivity (W/mK) | Mechanical strength (MPa) |
|---|---|---|---|
| Cu-OFE | 58–60 | 386–394 | 220–410 |
| CuBe2 | 6–8 | 105–120 | 1100–1380 |
| SS316LN | 1.4 | 14–17 | 485–690 |
| Ti6Al4V | 2.3 | 6.7 | 900–1120 |
| Nb | 6.3 | 53.7 | 275 |
| Ta | 7.7 | 57.5 | 180–240 |

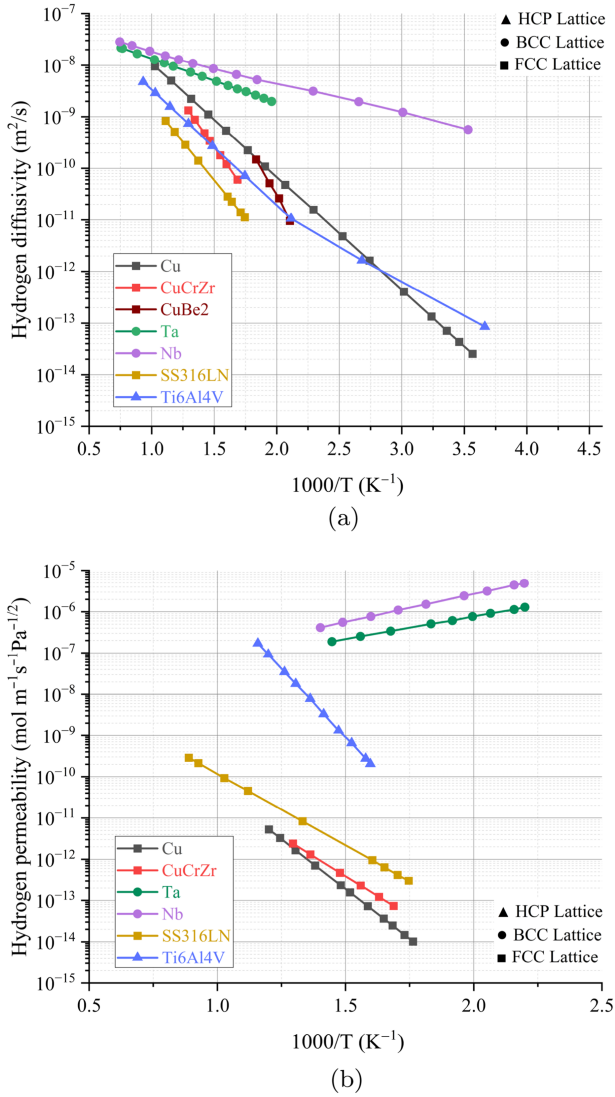

FIG. 1. (a) Compilation of hydrogen diffusion for different materials in function of reciprocal temperature, taken from different sources. Cu—[34], CuCr1Zr—[29], [25], CuBe2—[35], SS316LN—[29], Ti6Al4V—[36,37], Ta—[30], and Nb—[38], [30]. (b) Compilation of hydrogen permeability for different materials in function of reciprocal temperature, taken from different sources. Cu—[25], SS316LN—[29], CuCr1Zr—[29], [39], Ta—[40], Nb—[41], and Ti6Al4V—[42].

$$\phi = K_s D = D_0 S_0 \exp[-(E_d + E_s)/RT]. \quad (3)$$

As hydrogen from beam bombardment accumulates inside the metal lattice, if its amount exceeds the solubility limit, the internal pressure within the material can exceed the tensile strength of the metal, resulting in the formation of blisters in the material surface. Data extrapolated from different sources have been compiled together, as shown in Fig. 1, with (a) and (b) representing the hydrogen diffusion and permeability in a material's bulk in function of reciprocal temperature, respectively. Both coefficients for each material follow an Arrhenius-type equation. According to the literature, results for the same material can have substantial variability in terms of order of magnitude, as both diffusivity and permeability coefficients will vary from the purity of the material, contamination, and the measurement technique used in the experiments. Hydrogen diffusion is moderated by many factors, such as the size and accessibility of interstitial sites in the material, which differs from lattice to lattice, binding energy of H in the lattice, temperature, and impurities.

The copper alloys follow a similar behavior as Cu, which is expected as Cu is the main element of their composition. The hydrogen permeability of CuBe2 was not found in the literature, however, as just mentioned, this alloy should follow very closely the Cu permeability. Ti6Al4V also follows a similar behavior as pure Ti alpha based, more results can be found in [30–33].

As mentioned, Ta and Nb are among the chosen materials for testing, with the highest hydrogen diffusion capabilities. According to [43], Ta is one of the materials with the highest resistance to the formation of blisters. In [44], it is explained that because of good hydrogen dissolution, blisters are not observed. Reference [45] also reports a decrease in the activation energies for hydrogen diffusion in both niobium and tantalum at low temperatures. The small values for activation energies in Nb and Ta have been associated in part, with their body-centered cubic (bcc) lattice type, which favors the mobility of hydrogen atoms, as bcc crystal lattices have in general more interstitial spaces, when compared with a hexagonal close-packed (hcp) or a face-centered cubic (fcc) structures.

Copper has a relatively low-hydrogen diffusivity, which can cause the appearance of blisters in the surface due to the low rate of hydrogen atoms through the lattice. The additional low-hydrogen solubility consequently results in a very low-hydrogen permeability, even lower than austenitic steels. However, Cu has a higher diffusivity when compared with austenitic steels [46].

CuCr1Zr and CuBe2, copper based alloys, have similar diffusion characteristics when compared with pure copper. However, the addition of alloying elements can reduce the overall mobility of hydrogen ions by also introducing defects or trapping sites for hydrogen. According to [25], the permeability values for CuCr1Zr remain close to the ones in pure copper, but the addition of chromium has been associated with an increase in solubility and a decrease in the hydrogen transport through the lattice (diffusion). Both Cr and Zr, which act as strengtheners in copper, also can act as site traps for H atoms, making the trapped atoms less likely to contribute to the diffusion process.

CuBe2, being an alloy with approximately 98% Cu, the majority of the diffusion species should be copper [47], so in terms of diffusion, the behavior of CuBe2 should be similar. The small addition of beryllium has been studied in [48], which reports that the addition of Be to Cu causes the hydrogen permeability to decrease by a factor of 3 (relative to pure Cu) in the temperature range of 700 − 500 °C. Studies made in [49] demonstrated that at least five atoms of hydrogen can be accommodated in a monovacancy, indicating that vacancies in beryllium are efficient hydrogen traps. Additionally, the diffusion and solubility behavior of CuBe2 has been reported in [35], showing that in terms of diffusion, CuBe2 exhibits similar results when compared with pure oxygen-free copper and higher diffusion rate than the 300-series austenitic stainless steels.

For titanium, [32] reports that the presence of $\beta$ phase in Ti grade 5 alloy seems to be enough to dissolve hydrogen without producing either hydride precipitation or cracking. The vanadium addition in the alloy, with different concentrations, has been studied in [50]. It has been concluded that the diffusivity of hydrogen increases with the addition of V in the Ti base material. On the other hand, the presence of aluminum is associated with a decrease in diffusivity and solubility of hydrogen in the alloy [51,52]. Ti6Al4V has a $\alpha$ microstructure, with a hcp structure. This type of crystal lattice generally has less interstitial sites available for hydrogen diffusion when compared with bcc crystal lattices.

Austenitic stainless steels present high content of elements with a complex chemical composition. SS316LN presents a low diffusivity and high solubility of hydrogen due to its fcc structure [53]. Reference [54] explains that the alloy compositions do not influence the hydrogen permeability but affects the hydrogen diffusion process, i.e., diffusivity appears to depend on the alloy. The presence of Cr and Mo tends to reduce H diffusivity and to increase solubility; however, this effect is not as large as in bcc materials. In a study about hydrogen permeation in different austenitic steels, Ref. [55] explains that the hydrogen solubility correlates mainly with the main alloying elements, concluding that solubility is higher when Cr content increases and the content of Ni decreases. Also, N has been reported to increase solubility. More information about diffusion process, along with diffusion, solubility, and permeability parameters, in stainless steels can be found in Refs. [54–57].

In this paper, we will be focusing on the results of CuCr1Zr, Ta, CuBe2, and SS316LN, and only briefly summarize for completeness of the results of Cu-OFE, Ti6Al4V, and Nb which have been covered in [7].

### B. Experimental setups: Irradiation and LES system

The Large Electrode System (LES) is an experimental setup that allows the test of the samples under high-gradient electric field. The purpose of such a system comes from the need to replicate rf conditions as close as possible, to study breakdowns appearance in a way which is less expensive and less time consuming compared to rf test operations [58]. The LES enables the testing of one pair of cathode-anode electrodes using high-voltage short pulses. The vacuum chamber where the electrodes are inserted allows for an ultrahigh vacuum and is connected to a Marx generator that provides high-voltage pulses up to a repetition rate until 6 kHz. Measurements of current and voltage are taken from the connected oscilloscope. The main chamber has four apertures/windows which enables the external installation of cameras. Currently, two CCD cameras are installed in order to capture the light emitted from the individual breakdowns, software triangulation is used to extrapolate the exact position where the breakdown took place on the electrode surface and with the help of a data acquisition system create a mapping of the events location during conditioning. More information regarding the LES system, the breakdown localization technique, and details on the standard procedure that is followed for progressively increasing the electric field ("conditioning") can be found in [59]. For the present experiments, testing in the LES system is done by applying pulses of 100 μs and a repetition rate of 200 Hz to stay closer to the RFQ conditions while aiming at attaining a maximum breakdown rate of $10^{-5}$ per pulse as in the standard conditioning experiments.

The aim of manufacturing two pairs of cathode-anode for each material allowed us to use one to irradiate the cathode before submitting it to high-field pulsing tests in the LES chamber. It is commonly accepted that the cathode surface determines the development of breakdowns in the LES system [60]. The nonirradiated pairs are also tested, representing the reference results for the material in question. Comparison between the irradiated and nonirradiated pairs, in terms of performance in reaching an electric surface field, is studied.

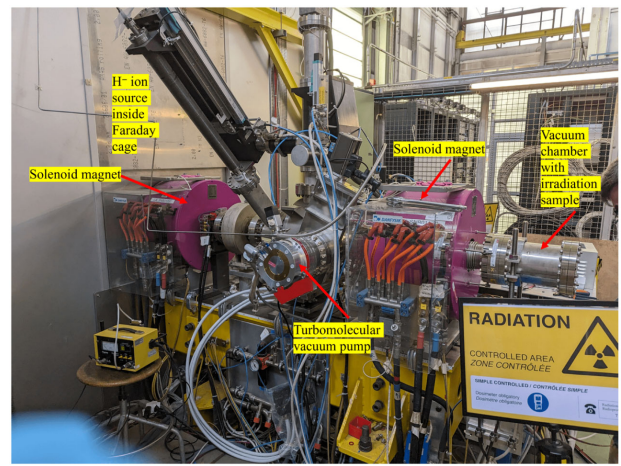

FIG. 2. Photograph of the irradiation setup at CERN, Geneva. The ion beam travels from the left to the right, having a chamber at the end dedicated for the mounting of one electrode.

Figure 2 shows the ion source irradiation facility used to irradiate all the samples. The electrodes are mounted using a dedicated chamber installed at the end of the structure. Specific details about the irradiation apparatus can be found in [3]. In the current RFQ at CERN, It has been reported that the maximum deposition of particles in the structure during operation is approximately $2 \times 10^{11}$ p/s/mm$^2$ [3]. Taking into consideration this value and the beam loss profile along the RFQ, parameters for the irradiation of the samples under study were chosen. Low-energy H$^-$ beam at 45 keV, corresponding to the energy of the beam entering the RFQ from the ion source, was used to irradiate one cathode of each material, each one with a duration of approximately 50 h. The pulse duration, repetition rate, and peak current were 600 μm, 0.83 Hz, and 20 mA, respectively. This irradiation resulted in a deposition of $1.2 \times 10^{19}$ H$^-$ ions in each target over the entire irradiation, which represents approximately 10 days of beam losses during RFQ operation.

### C. Microscopic techniques

SEM observations were the main techniques used to have a comprehensive inspection of the surface status in all electrodes. After manufacturing and metrology, SEM observations were done to validate the final surface, before installing electrodes for testing.

Due to the profiling of blistering, breakdowns, and surface roughness under study, secondary electron detector in the SEM was the most used to study the topography of our samples. This detector is installed at a 45° from the vertical axis of the beam gun of the SEM, allowing for more topographical features to be detected. Apart from SE2, inLens detector, which is more sensitive to low-energy electrons and installed in the same vertical axis as the beam, was used when a more superficial analysis was needed. Finally, back-scattered electron (BSE) detector, which detects variations in atomic number, leading to different intensities of the signal, was used to identify the presence of different elements apart from the sample's material. Different elements are represented with different levels of brightness, showing the lighter elements with a darker contrast. As our electrodes have a superficial area of approximately 2800 mm$^2$, using BSE detector proved to be very useful in scanning large areas of the surface. All the SEM analysis was complemented with energy-dispersive x-ray spectroscopy (EDS). This chemical analysis technique allowed us to identify qualitatively external elements that may be present in the surface of the electrodes.

### III. RESULTS AND DISCUSSION

### A. Conditioning results

Table III shows the results from all the materials tested in this project, and for each material, the maximum and stable field that irradiated and nonirradiated electrodes have been able to achieve. In most of these experiments, the conditioning process has been performed manually, where small increments of voltage are applied only after the previous target field is reached and stable.

TABLE III. Summary table specifying the maximum and stable field reached by all the tested materials, including results from [7].

| Material | Irradiation | Maximum field (MV/m) | Stable field (MV/m) |
|---|---|---|---|
| SS316LN | No | 120 | 120 |
| | Yes | 62.5 | 60 |
| CuBe2 | No | 110 | 90 |
| | Yes | 45 | 16.7 |
| CuCr1Zr | No | 87.5 | 82.5 |
| | Yes | 29 | 25 |
| Ta | No | 60 | 60 |
| | Yes | 38.1 | 24 |
| Ti6Al4V | No | 110 | 100 |
| | Yes | 95 | 90 |
| Cu-OFE | No | 80 | 80 |
| | Yes | 80 | 80 |
| Nb | No | 94 | 80 |
| | Yes | 42 | 21.7 |

The following plots show the conditioning results for copper beryllium, copper chromium zirconium, stainless steel, and tantalum. In each plot, four different curves can be seen. The extent of the entire conditioning process can be read on the horizontal axis, as the cumulative number of pulses. The blue curve represents the applied electric field during the conditioning process for the nonirradiated material, while the magenta curve has the same representation for the irradiated material. These two curves are composed with dots—each dot shows the current electric field which is registered at every 100 000 pulses. The electric field for these curves can be read in the left vertical axis, calculated from the ratio between the voltage applied and the physical gap in use (60 μm in this cases). When observing the curve, sudden drops in electric field can be seen. These drops represent a breakdown or a multiple breakdowns event on the material surface. After a breakdown happens, the recovery of voltage is set, corresponding to the same level of voltage prior to the breakdown, and the system automatically increases voltage back to this target—for more information regarding the voltage behavior upon a vacuum arc, consult Ref. [59]. The cumulative breakdown events are counted during the process and can be read through the right vertical axis. The curves in orange and green represent the cumulative number of breakdowns (No. BD) during the conditioning tests for the nonirradiated and irradiated materials, respectively.

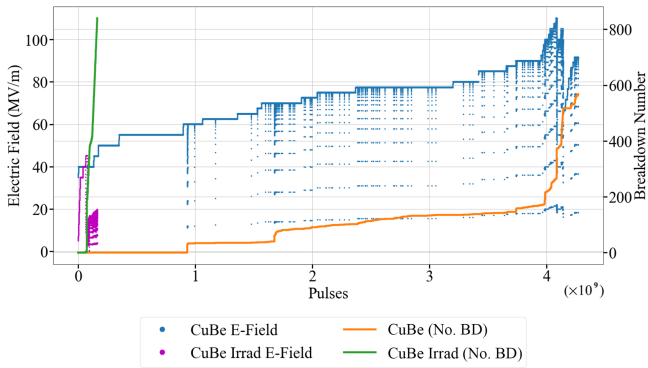

FIG. 3. Conditioning of the irradiated and nonirradiated pairs of CuBe2 electrodes.

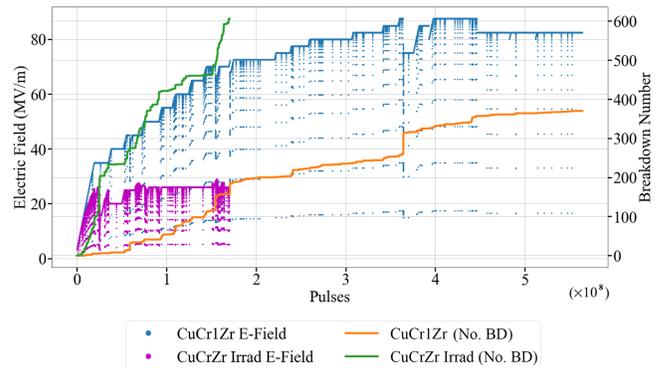

FIG. 4. Conditioning of the irradiated and nonirradiated pairs of CuCr1Zr electrodes.

### 1. CuBe2

Figure 3 shows the conditioning for CuBe2 electrodes irradiated and nonirradiated.

For the nonirradiated electrodes, a maximum field of 110 MV/m was reached. However, at high fields, breakdowns started to cluster at a specific location, where it appeared to be impossible for the field to remain stable. Taking into consideration that the stable field was directly affected by this cluster of breakdowns, a retest on new CuBe2 nonirradiated electrodes will be done in the future to assess whether the maximum and stable field can achieve higher results. In contrast with the results above, the irradiated pair has only reached a maximum field of 40 MV/m after approximately 100 million pulses as seen by the magenta curve in Fig. 3. The high number of breakdowns (green curve) at that field resulted in a sudden drop in voltage, leading to an unrecoverable condition. The irradiated electrodes did not recover back to the earlier maximum voltage, instead, the stable field ended up at approximately 17 MV/m.

### 2. CuCr1Zr

Nonirradiated CuCr1Zr has reached a maximum field of 87.5 MV/m, as shown in Fig. 4. After reaching the maximum value, the electrodes experienced a sudden decrease in the field to 75 MV/m, and consequently, a big increase in the breakdown rate. After this, a retry of stepping up the voltage was done. The electrodes were able to reach the maximum field again and hold it for a longer period but ended up stabilizing at 82.5 MV/m. For the irradiated CuCr1Zr, this pair of electrodes has reached a maximum stable field of 26 MV/m. There was not any breakdown cluster registered during conditioning or observed in the surface analysis; however, the majority of the breakdowns were concentrated in the irradiated area.

The deposition of particles from the irradiation is visible in Fig. 6(b) in the darkest circle region.

### 3. SS316LN

Figure 5 shows the conditioning for the irradiated and nonirradiated SS316LN electrodes. Stainless Steel nonirradiated electrodes have shown very good results in terms of conditioning, being the material that has reached the highest electric field, among all the tested materials.

During conditioning, the high voltage cable between the Marx generated and the connection to the chamber broke 2 times, due to the maximum voltage supported by the cable. In Fig. 5, this can be observed in the two reconditioning events at approximately 0.65 and 0.9 billion pulses in the blue dots. After replacing the cable with one that could hold the applied voltages, the electrodes were able to reach a maximum field of 120 MV/m. On the other hand, the irradiated electrodes only reached a stable field of 60 MV/m. When trying to increase the voltage applied, the electrodes reach the maximum field of 62.5 MV/m, but without being able to stabilize in that field and resulting in a significant increase in the number of breakdowns. From Fig. 6(c), it can be seen that the large increase in breakdowns is locally coincident with the irradiated region. The irradiated region corresponds to the darkest half-circle seen on the left part of the electrode surface.

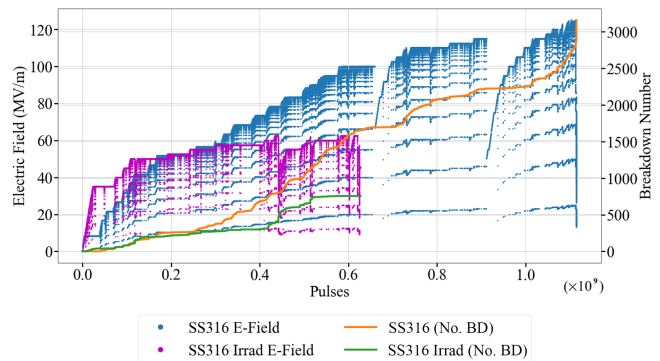

FIG. 5. Conditioning of the irradiated and nonirradiated pairs of SS316LN electrodes.

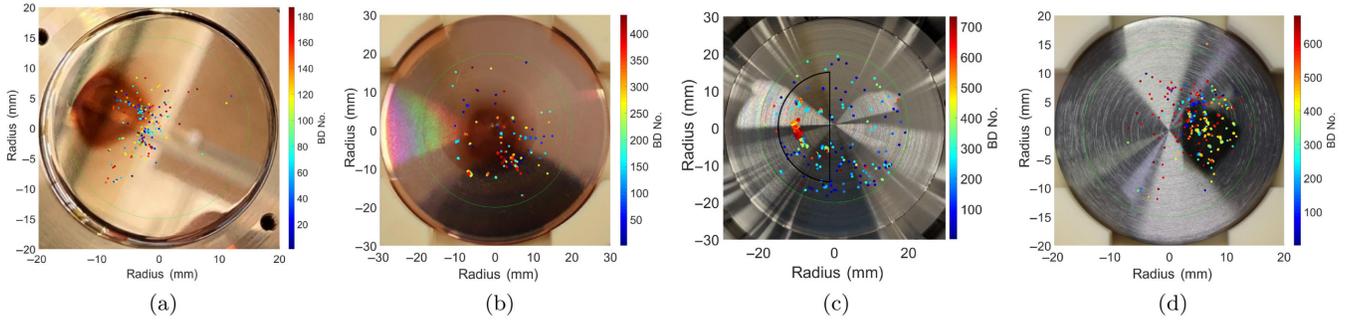

FIG. 6. Images taken before closing the chamber of the different irradiated cathodes overlaid with the breakdown locations after conditioning. Breakdowns are colored from dark blue to dark red as the first to last breakdown. (a) CuBe2, (b) CuCr1Zr, (c) SS316LN, (d) Ta. The green circles in all figures represent the coincident surface diameter of the anode electrode. The irradiated targeted regions can bee seen as a darker color on the electrode surfaces. For the case of SS316LN, the irradiated region is not well distinguished and it has been marked with a black half-circle.

#### 4. Ta

Figure 7 shows the conditioning of the irradiated and nonirradiated Ta electrodes. For the nonirradiated, the maximum and stable field reached was 60 MV/m. For the irradiated pair, the maximum field reached was 35 MV/m; however, after just $10^7$ pulses, the field dropped to around 23 MV/m and was unable to recover from that value. After the conditioning, no cluster was observed on the surface of the electrodes; however, the effects of irradiation are very clear, showing the concentration of breakdowns mainly in the impact zone of the H$^-$ beam. Figure 6(d) shows the locations of the breakdowns in the irradiated cathode, in which it can be observed that a larger quantity of breakdowns are concentrated in the irradiation zone.

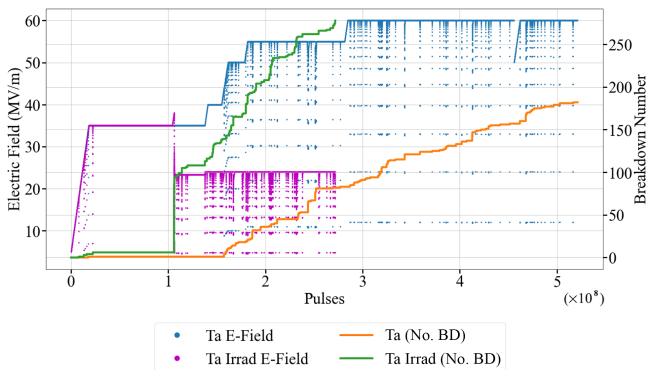

FIG. 7. Conditioning of the irradiated and nonirradiated pairs of Ta electrodes.

### B. Microscopic observations and discussion

For all the nonirradiated materials, microscopic analysis has shown that the breakdowns from the LES tests are uniformly distributed throughout the surface of the electrodes. Also, breakdowns shapes and size are in accordance with the results reported in [7]. SEM observations confirmed blistering on the surface coincident with the irradiated region. For the copper based materials, i.e., CuBe2 and CuCr1Zr blistering caused by the poor hydrogen diffusion in the material is visible, as was the case for pure Cu-OFE. Blistering on the copper electrodes, as previously studied, can be seen in [7]. The beam density distribution follows a Gaussian distribution which results in more particles arriving at the center of the targeted region compared with the outer regions (for copper based materials in which blisters are observed, the nonuniformity of beam density is well observed by the different density of blisters). The distribution of the energy profile in all irradiation is uniform in the entire targeted region, which means that each ion reaches the surface with 45 keV. Copper was the material in which studies have been made to understand the penetration depth and the effects of irradiation in terms of blistering, where it was concluded that the penetration depth for the H ions (approximately between 250 and 300 nm) is independent of the beam density distribution. More information regarding the blistering effects in copper can be found in [61].

As expected, the good diffusion of hydrogen in Ta is confirmed with no blistering appearing on the surface after irradiation, as was also previously observed in Nb and Ti6Al4V. The same behavior happens for SS316LN; however, the nonappearance of blisters cannot be attributed to its hydrogen diffusion coefficient of the material but rather to the high-strength characteristics of this material [62].

For CuCr1Zr, it is observed that the blisters seem to follow the machining marks of the surface, as shown in Fig. 8(a). This effect is more visible in the center of the electrode surface, where the machining grooves are more pronounced. Figures 8(b) and 8(c) shows the surface of the same electrode, after conditioning test. In Fig. 8(b), it can be seen a different contract in surface coloring: the whiter region has a more rough surface where the multiple breakdowns are located. The more magnified Fig. 8(c) shows the contours of a carbon layer surrounding the edges of the small breakdowns. More detailed findings regarding the carbon effects are discussed below.

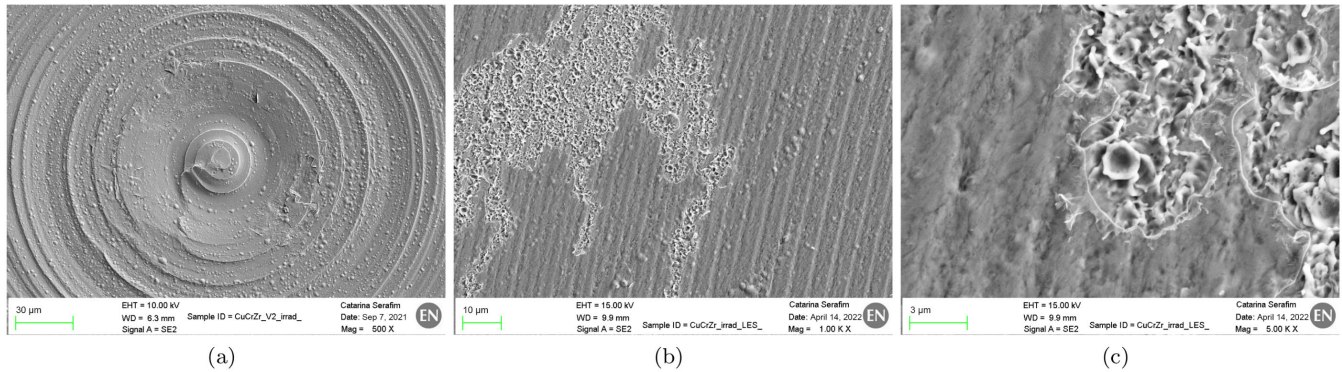

FIG. 8. SEM imaging, using secondary electron detector, of (a) center of CuCr1Zr electrode's surface after H⁻ beam irradiation and before LES testing. (b), (c) breakdowns in CuCr1Zr after H⁻ beam irradiation and LES testing.

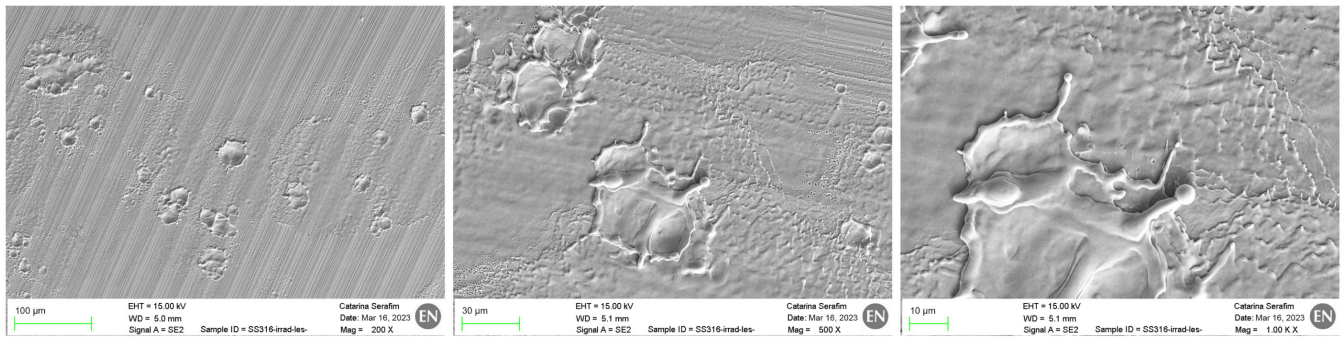

FIG. 9. Breakdown images, using secondary electron detector from SEM with different magnification, of SS316LN electrode after H⁻ beam irradiation and LES testing.

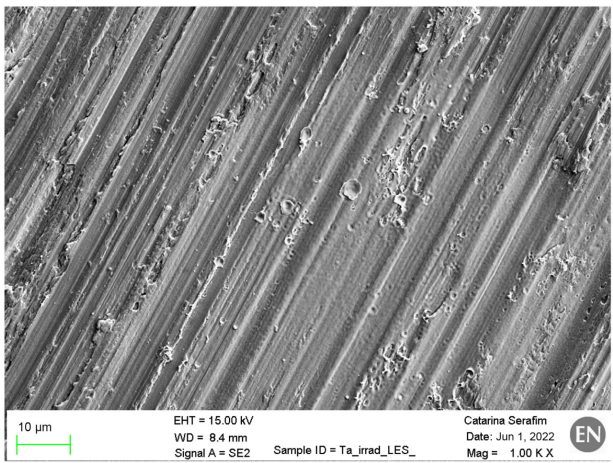

FIG. 10. Breakdown image, acquired using SEM secondary electron detector, on the tantalum electrode after H⁻ beam irradiation and LES testing

In all the irradiated materials, breakdowns have dimensions between 1 and 50 μm, in the irradiated regions, proportional to the electric surface field that was achieved, see [63] for comparison. Figures 9 and 10 show the breakdowns inside the irradiated region of SS316LN and Ta electrodes, respectively. The sequential vacuum arcs and the spots temperature rise lead to local melting, forming craters on the material surface, typically with splashes of molten material in its surroundings. This behavior has been reported before in the specific case of stainless steel in [64]. The irradiated SS316LN was the one with a better response to irradiation of all the materials described in this section, not only reaching the highest electric surface field (60 MV/m) but also for being able to hold that same field without suddenly dropping in voltage. Also, stainless steel was the only material in which breakdowns dispersed more from the irradiated region. Nevertheless, such good response, does not exclude the main fact that the 60 MV/m reached by the irradiated SS316LN corresponds to half of the electric field reached by the nonirradiated SS316LN.

In all irradiated materials, the problem in reaching a stable field is predominant. The big difference between the values of surface electric field between the irradiated and nonirradiated pairs can only mean that irradiation has a direct cause in decreasing the conditioning process. The cause does not seem to be correlated with the blistering effect, as blisters are not observed in Ta or SS316LN. Based on surface characterization of all the materials made before LES testing, no defects on the surface were found that could explain the poor performance. Cu-OFE results covered in [7] are another indication of the nonrelation of the blistering effect with the breakdowns appearance. Chemical analysis was performed in the different stages of all testing, leading to the conclusion that a carbon contamination happened after irradiation procedure. EDS was performed at multiple points of the surface inside and outside the irradiated regions of the different cathodes. From this, it was observed that a higher amount of C coincident with regions where we have more deposition of $H^-$ particles, and the C percentage diminishes as we go further away from the irradiated region. Values ranged between 5% and 12% in weight, in contrast with the nonirradiated regions in which only 1%–2% was observed.

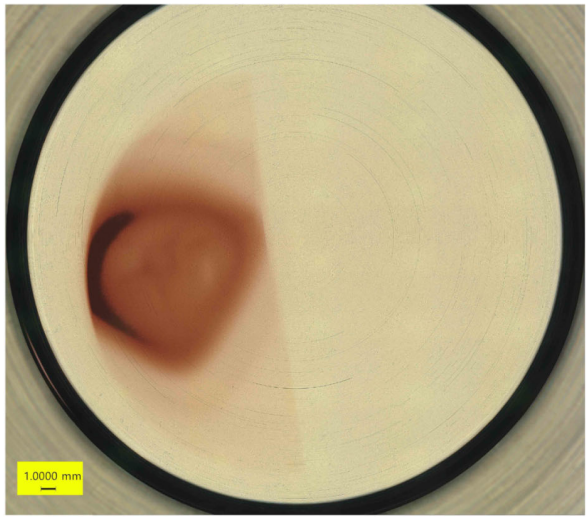

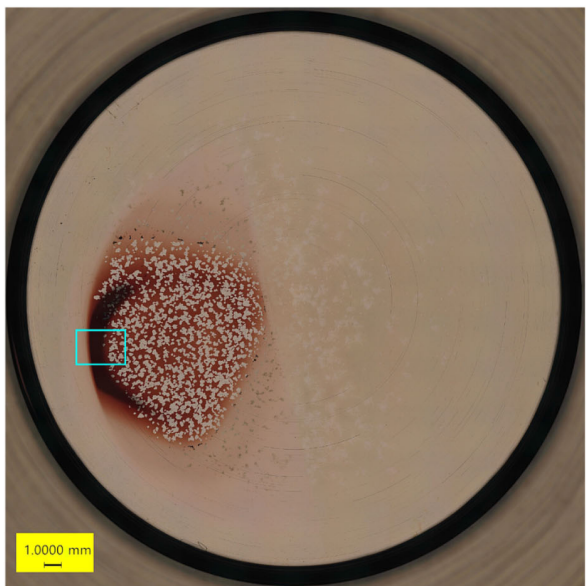

FIG. 11. Optical microscopy imaging of (a) CuBe2 cathode after being exposed with irradiation and before conditioning test; (b) CuBe2 cathode after being exposed with irradiation and after conditioning test. Blue square indicates the location of Fig. 12 on the surface of the electrode.

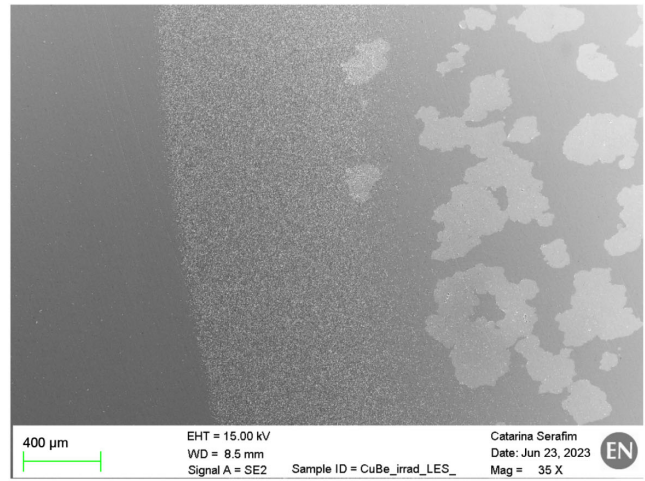

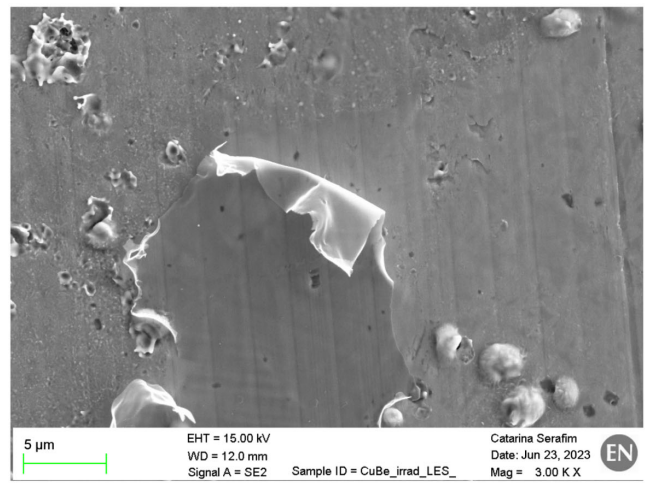

FIG. 12. (a) SEM imaging using secondary electron detector, with 35× of magnification, of CuBe2 cathode after irradiation and LES test. (b) SEM imaging using secondary electron detector, with 3000× of magnification, of CuBe2 cathode after irradiation and LES test, showing a layer of C partially removed from the surface.

The cause of the carbon presence in all the irradiated surfaces might be attributed to contamination from the irradiation process, possibly from ion-induced cracking of residual hydrocarbon gases onto the surfaces. The presence of this superficial carbon has provided unusual conditioning behaviors in LES tests, not yet seen until the current date.

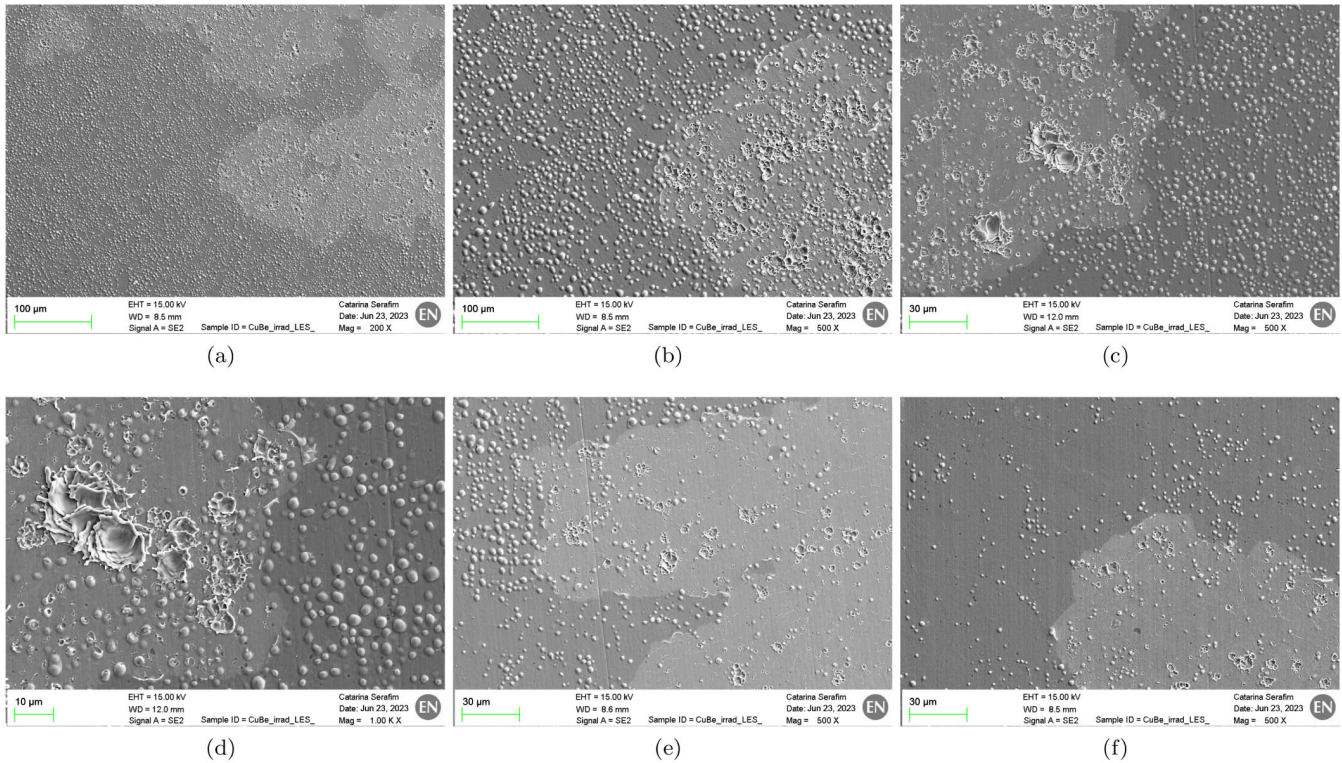

FIG. 13. (a)–(d) SEM imaging, using secondary electron detector, of CuBe2 cathode after irradiation and LES test. The regions in focus were taken where there was the highest deposition of particles from irradiation. (e),(f) SEM imaging, using secondary electron detector, of CuBe2 cathode after irradiation and LES test. The regions in focus were taken in the transition zone.

In Ta and SS316LN materials, when using SEM imaging, the presence of a carbon layer goes almost unnoticed. However, for CuBe2 and CuCr1Zr, the layer of carbon that has been formed on the surface is very visible. Below, we share images of CuBe2 cathode after irradiation and conditioning test using the LES system.

Figure 11(a) shows an optical image of the surface of the cathode electrode after it has been exposed to irradiation. The irradiated region is shown with a darker coloration, forming a half-circle shape, coincident with the gasket used during irradiation to limit the irradiated spot in the target. The darkest C-shape is where we had the highest deposition of particles, and it is where the highest density of blisters is found. After irradiation, LES tests were performed with this electrode. Figure 11(b) shows the state of the surface after the test, where the appearance of white spots covering the irradiation zone and its surroundings was observed.

Further analysis, using SEM and EDS, confirmed that the white regions represent the regions where carbon has been removed from the surface, designated as exfoliated regions. In SEM imaging, exfoliated regions appeared with a lighter contrast compared with the rest of the darker surface, as shown in Fig. 12(a).

The carbon layer is a fragile and thin layer, measured to have approximately 60 nm of thickness in the regions with higher ion deposition [61]. Figure 12(b) shows the carbon layer peeling out from the surface (machining marks of the electrode surface are visible beneath the C layer). It becomes difficult to access the origin for the removal of the carbon. Analysis made postirradiation and pre-LES testing showed the carbon layer contamination without signs of degradation or cracking, which proves that the different regions where carbon has been removed appeared during the high voltage tests. At this stage, it is hypothesized that the initial stress from a vacuum arc or high-voltage exposure could have initiated a crack on the C layer, from which it triggered a peeling effect enhanced by the vacuum environment. Also, outgassing coming from the electrode could originate additional stress against the C layer, provoking cracking and peeling.

Additionally, it has been observed that breakdowns, from the high pulsing test, are only concentrated inside these exfoliated areas. This result is common for all the different regions: either in regions where there was more deposition of particles and therefore more density of blisters per area [Figs. 13(a)–13(d)], or in transition zones, where there is a decrease in blister's density as we move further from the high-density zone [Figs. 13(e)–13(f)]. This result may bring further insights to the scientific community of vacuum breakdowns studies, since it has not been yet fully understood the mechanisms that triggers the appearance of a breakdown. From the current results, we can summarize the following observations: (i) in an irradiated sample that has been contaminated with carbon, the number of breakdowns in the irradiated region is much bigger than the number of breakdowns in the nonirradiated region; (ii) in post-LES testing observations, breakdown sites were not observed in regions that still had a carbon contamination layer, and regions where breakdown sites were observed did not have a carbon contamination layer—the carbon removal happened during high pulsing field test; (iii) the final electric field achieved in electrodes contaminated with carbon is highly affected by its presence, when compared with the performance of electrodes that have not been irradiated and free from carbon contamination; (iv) the presence of blistering on the surface does not seem to have a correlation with the breakdown triggering, nor an effect on the maximum field reached.

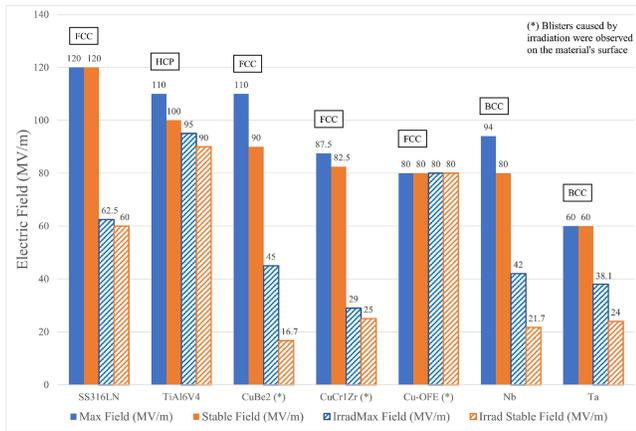

FIG. 14. Bar plot showing the results for all the tested materials in the LES system. The solid bars represent the results of nonirradiated electrodes. The lined bars represent the results for the irradiated electrodes. The blue and orange colors represent, respectively, the maximum and stable electric field achieved during testing.

Accounting for the results presented here and in [7], the final summary of all the materials in terms of surface electric field, when tested with LES system, was compiled together in Fig. 14, showing the results from all materials in order of best stable field achieved.

## IV. CONCLUSIONS

We have measured the maximum field attained by several materials in the LES pulsed dc high-voltage system at a constant breakdown rate. Taking into consideration only the tests of nonirradiated materials, SS316LN shows to be the best material in reaching the highest and stable surface electric field (120 MV/m). Of the copper alloys, CuBe2 attains the highest performance reaching 90 MV/m. As a reference, standard Cu-OFE electrodes typically reach about 80 MV/m. As the main purpose of this study is to find a better substitute material for the manufacturing of an RFQ, the interest comes to materials that can perform better than copper, which is the material traditionally used.

All of the irradiated materials have presented a big decrease in performance when compared with the non-irradiated electrodes. Blistering, as shown in copper alloys, does not seem to be directly correlated with this reduction. The reduced performance is instead associated with a carbon layer observed on the cathode's surface after irradiation. This may require further investigation regarding the role of carbon in the breakdown studies. For future testing, ways to remove the carbon layer from the electrodes surface are being investigated, which is not simply feasible with standard cleaning processes used in accelerator vacuum technology such as degreasing.

## ACKNOWLEDGMENTS

The authors thank all the members involved in the WP10 of the LINAC4 spare RFQ project. Also, all the LINAC personnel that have made possible the irradiation campaigns, during the last 3 years, for all the materials. Additionally, a special acknowledgment for the MME team at CERN which provided access to the microscopy facilities that made possible all the analysis and for the members of the workshop team, responsible for the machining and metrology of the electrodes.

The authors declare that the research was conducted in the absence of any commercial or financial relationship that could be construed as a potential conflict of interest.

[1] A. Lambert *et al.*, High-intensity proton RFQ accelerator fabrication status for PXIE, in *Proceedings of the 6th International Particle Accelerator Conference, Richmond, VA* (JACoW, Geneva, Switzerland, 2015), WEPTY045.

[2] Z. Zhang, X. Xu, Y. He, S. Zhang, C. Wang, S. Zhang, C. Li, and Y. Huang, Design of a radio frequency quadrupole for a high intensity heavy-ion accelerator facility, Phys. Rev. Accel. Beams **25**, 080102 (2022).

[3] S. Calatroni, Low-energy h- irradiation of accelerators components: Influence of material and material properties on the breakdown resistance at high electric fields, in *Proceedings of the 9th International Workshop on Mechanisms of Vacuum Arcs (MeVArc 2021)* (2021), https://indico.cern.ch/event/966437/timetable/?view=standard_numbered.

[4] A. Grudiev, S. Calatroni, and W. Wuensch, New local field quantity describing the high gradient limit of accelerating structures, Phys. Rev. ST Accel. Beams **12**, 102001 (2009).

[5] W. Wang, J. Roth, S. Lindig, and C. H. Wu, Blister formation of tungsten due to ion bombardment, J. Nucl. Mater. **299**, 124 (2001).

[6] A. Descoeudres, F. Djurabekova, and K. Nordlund, DC breakdown experiments with cobalt electrodes, CERN, Switzerland, Report No. CLIC-Note-875, CERN-OPEN-2011-029, 2010.

EFFECTS OF H⁻ LOW BEAM IRRADIATION AND … PHYS. REV. ACCEL. BEAMS **28,** 013101 (2025)

[7] C. Serafim, R. Peacock, S. Calatroni, F. Djurabekova, A. T. P. Fontenla, W. Wuensch, S. Sgobba, A. Grudiev, A. Lombardi, E. Sargsyan, S. Ramberger, and G. Bellodi, Investigation on different materials after pulsed high field conditioning and low-energy H$^-$ irradiation, Front. Phys. **12**, 3 (2024).

[8] P. Hanzelka, V. Musilova, T. Kralik, and J. Vonka, Thermal conductivity of a CuCrZr alloy from 5 K to room temperatures, Cryogenics **50**, 737 (2010).

[9] R. Li, E. Guo, Z. Chen, H. Kang, W. Wang, C. Zou, T. Li, and T. Wang, Optimization of the balance between high strength and high electrical conductivity in CuCrZr alloys through two-step cryorolling and aging, J. Alloys Compd. **771**, 1044 (2019).

[10] M. Corporation, *Guide to High Performance Alloys* (Mayfield Heights, OH, 2019), https://hkvxni.files.cmp.optimizely.com/download/bb4cc744a04211ee9c2cea13b3482b31.

[11] I. Savchenko and S. Stankus, Thermal conductivity and thermal diffusivity of tantalum in the temperature range from 293 to 1800 K, Thermophys. Aeromech. **15**, 679 (2008).

[12] C. Y. Ho, R. W. Powell, and P. E. Liley, Thermal conductivity of the elements, J. Phys. Chem. Ref. Data **1**, 279 (1972).

[13] M. Fouaidy, Thermal conductivity of niobium and thermally sprayed copper at cryogenic temperature, IOP Conf. Ser. **1241**, 012012 (2022).

[14] A. J. Fletcher, *Thermal Stress and Strain Generation in Heat Treatment* (Springer Netherlands, 1989), https://api.semanticscholar.org/CorpusID:136482545.

[15] T. K. Chu and C. Y. Ho, *Thermal Conductivity and Electrical Resistivity of Eight Selected AISI Stainless Steels* (Springer US, Boston, MA, 1978).

[16] X. Tan, P. P. Conway, and F. Sarwar, Thermo-mechanical properties and regression models of alloys: AISI 305, CK 60, CuBe2 and Laiton MS 63, J. Mater. Process. Technol., **168**, 152 (2005).

[17] K. Zhang, E. Gaganidze, and M. Gorley, Development of the material property handbook and database of CuCrZr, Fusion Eng. Des. **144**, 148 (2019).

[18] F. Han, Y. Jiang, F. Cao, L. Cai, and S. Liang, A hard/soft layered micro-structured CuCrZr alloy with strength, ductility and electrical conductivity synergy, Mater. Charact. **199**, 112836 (2023).

[19] Available at https://www.kme-archiv.com/fileadmin/DOWNLOADCENTER/COPPER%20DIVISON/4%20Industrial%20Rolled/3_Datasheets/3_Datasheets_new_2019/Cu-OFE_04_2019.pdf [Accessed on 21/04/2024].

[20] Available at https://www.metalcor.de/en/datenblatt/137/ [Accessed on 21/04/2024].

[21] Available at https://www.metalcor.de/en/datenblatt/133/ [Accessed on 21/04/2024].

[22] Available at https://www.spacematdb.com/spacemat/manudatasheets/TITANIUM%20ALLOY%20GUIDE.pdf [Accessed on 21/04/2024].

[23] B. Bonin, *Materials for Superconductive Cavities*, Commissariat à l'Energie Atomique, DSM/DAPNIA, France, CAS - CERN Accelerator School: Superconductivity in Particle Accelerators, 1996, pp. 191–200, https://cds.cern.ch/record/399568.

[24] Available at https://www.finetubes.co.uk/-/media/ametek-finetubes/files/products/materials/fine_tubes_-_alloy-316ln.pdf?la=en&revision=11cba9fc-9f74-4e25-af17-a65a0507ece6&hash=1BBE7F8485179B91D498D11F72D92ED1&utm_source=https://www.google.com/&utm_medium=undefined [Accessed on 21/04/2024].

[25] I. Penalva, G. Alberro, F. Legarda, G. Esteban, and B. Riccardi, *Interaction of Copper Alloys with Hydrogen* (IntechOpen, Spain, 2012), ISBN 978-953-51-0160-4, 10.5772/34469.

[26] B. W. E. Materials, *Guide to Copper Beryllium* (Cleveland, OH, 2002), https://www.matthey.ch/fileadmin/user_upload/downloads/Fichiers_PDF/GuideToCopperBeryllium_EN.pdf.

[27] N. B. Eurpoe, Beryllium copper alloys, technical guide (2018), https://www.ngk-alloys.com/NGK_Berylco_Design_Guide_En.pdf.

[28] D. Terentyev, M. Klimenkov, and L. Malerba, Confinement of motion of interstitial clusters and dislocation loops in bcc Fe-Cr alloys, J. Nucl. Mater. **393**, 30 (2009).

[29] T. A. Shishkova, A. V. Golubeva, and M. B. Rozenkevich, Isotope effect in the interaction between hydrogen and fusion reactor materials, Russ. J. Phys. Chem. A **97**, 2079 (2023).

[30] A. Turnbull, Hydrogen diffusion and trapping in metals, in *Gaseous Hydrogen Embrittlement of Materials in Energy Technologies*, edited by P. Gangloff and B. P. Somerday, Woodhead Publishing Series in Metals and Surface Engineering Vol. 1 (Woodhead Publishing, United Kingdom, 2012), pp. 89–128, 10.1533/9780857095374.1.89.

[31] O. Palumbo, F. Trequattrini, S. Tosti, A. Santucci, and A. Paolone, Hydrogen and deuterium solubility, diffusivity and permeability from sorption measurements in the Ni33Ti39Nb28 alloy, Molecules **28**, 1082 (2023).

[32] V. Madina and I. Azkarate, Compatibility of materials with hydrogen. particular case: Hydrogen embrittlement of titanium alloys, Int. J. Hydrogen Energy **34**, 5976 (2009).

[33] E. Tal Gutelmacher and D. Eliezer, The hydrogen embrittlement of titanium-based alloys, JOM **57**, 46 (2005).

[34] H. Magnusson and K. Frisk, Diffusion, permeation and solubility of hydrogen in copper, J. Phase Equilib. Diffus. **38**, 65 (2017).

[35] J. Yamabe, D. Takagoshi, H. Matsunaga, S. Matsuoka, T. Ishikawa, and T. Ichigi, High-strength copper-based alloy with excellent resistance to hydrogen embrittlement, Int. J. Hydrogen Energy **41**, 15089 (2016).

[36] N. L. Guiles and K. Ono, The effect of hydrogen o internal friction of several titanium alloys, U.S. Department of Energy Office of Scientific and Technical Information, Materials Department: School of Engineering and Applied Science, University of California, Los Angeles, US, 1975, https://www.osti.gov/biblio/7348076.

[37] H. Jehn, H. Specka, E. Fromm, and G. Horz, Physics data: Gases and carbon in metals, thermodynamics, kinetics and properties. pt.v: group IV a metals (1) titanium (Ti), Energy Physik Mathematik GmbH Karlsruhe (1979), https://www.osti.gov/etdeweb/biblio/6792108.

[38] J. Völkl and G. Alefeld, in *Diffusion of Hydrogen in Metals* (Springer Berlin Heidelberg, Berlin, Heidelberg, 1978), pp. 321–348.

[39] E. Serra and A. Perujo, Hydrogen and deuterium transport and inventory parameters in a Cu-0.65Cr-0.08Zr alloy for fusion reactor applications, J. Nucl. Mater. 258-263, 1028 (1998).

[40] H. Yukawa, T. Nambu, and Y. Matsumoto, 13 - Design of group 5 metal-based alloy membranes with high hydrogen permeability and strong resistance to hydrogen embrittlement, in *Advances in Hydrogen Production, Storage and Distribution*, edited by A. Basile and A. Lulianelli (Woodhead Publishing, United Kingdom, 2014), pp. 341–367, 10.1533/9780857097736.3.341.

[41] J. Conde, M. Marono, and J. Sánchez, Pd-based membranes for hydrogen separation: Review of alloying elements and their influence on membrane properties, Sep. Purif. Rev. 46, 152 (2016).

[42] V. Maroni and E. Van Deventer, Materials considerations in tritium handling systems, J. Nucl. Mater. 85-86, 257 (1979).

[43] V. T. Astrelin, A. V. Burdakov, P. V. Bykov, I. A. Ivanov, A. A. Ivanov, Y. Jongen, S. G. Konstantinov, A. M. Kudryavtsev, K. N. Kuklin, K. I. Mekler, S. V. Polosatkin, V. V. Postupaev, A. F. Rovenskikh, S. L. Sinitskiy, and E. R. Zubairov, Blistering of the selected materials irradiated by intense 200 keV proton beam, J. Nucl. Mater. 396, 43 (2010).

[44] A. Badrutdinov, T. Bykov, S. Gromilov, Y. Higashi, D. Kasatov, I. Kolesnikov, A. Koshkarev, A. Makarov, T. Miyazawa, I. Shchudlo, E. Sokolova, H. Sugawara, and S. Taskaev, In situ observations of blistering of a metal irradiated with 2-MeV protons, Metals 7, 558 (2017).

[45] H. Wipf, *Diffusion of Hydrogen in Metals* (Springer Berlin Heidelberg, Berlin, Heidelberg, 1997), pp. 51–91.

[46] C. S. Marchi, Technical reference on hydrogen compatibility of materials. Copper alloys: Pure copper (code 4001), Sandia National Laboratories, Report No. SAND2012-7321. 2012.

[47] D. Eckman, B. Z. Rosenblum, and C. Q. Bowles, Diffusion bonding of beryllium-copper alloys, J. Mater. Sci. 27, 49 (1992).

[48] D. B. Butrymowicz, J. R. Manning, and M. E. Read, Diffusion in copper and copper alloys. Part III. Diffusion in systems involving elements of the groups IA, IIA, IIIB, IVB, VB, VIB, and VIIB, J. Phys. Chem. Ref. Data 4, 177 (1975).

[49] M. G. Ganchenkova, V. A. Borodin, and R. M. Nieminen, Hydrogen in beryllium: Solubility, transport, and trapping, Phys. Rev. B 79, 134101 (2009).

[50] H. Christ, S. Schroers, and F. Santos, Diffusion of hydrogen in titanium-vanadium alloys, Defect Diffus. Forum 237–240, 340 (2005).

[51] J. W. Wang and H. Gong, Adsorption and diffusion of hydrogen on Ti, Al, and TiAl surfaces, Int. J. Hydrogen Energy 39, 6068 (2014).

[52] D. Connétable, Theoretical study on hydrogen solubility and diffusivity in the γ-TiAl L10 structure, Int. J. Hydrogen Energy 44, 12215 (2019).

[53] E. Herms, J. Olive, and M. Puiggali, Hydrogen embrittlement of 316L type stainless steel, Mater. Sci. Eng. A 272, 279 (1999).

[54] S. Xiukui, X. Jian, and L. Yiyi, Hydrogen permeation behaviour in austenitic stainless steels, Mater. Sci. Eng. A 114, 179 (1989).

[55] Y. Chen, D. M. Santos, and C. A. Sequeira, Hydrogen diffusion in austenitic stainless steels, in *Diffusion in Solids and Liquids, DSL-2006*, Defect and Diffusion Forum Vol. 258 (Trans Tech Publications Ltd, Switzerland, 2007), pp. 322–326, 10.4028/www.scientific.net/DDF.258-260.322.

[56] C. San Marchi, B. Somerday, and S. Robinson, Permeability, solubility and diffusivity of hydrogen isotopes in stainless steels at high gas pressures, Int. J. Hydrogen Energy 32, 100 (2007).

[57] K. Forcey, D. Ross, J. Simpson, and D. Evans, Hydrogen transport and solubility in 316L and 1.4914 steels for fusion reactor applications, J. Nucl. Mater. 160, 117 (1988).

[58] A. Saressalo, I. Profatilova, A. Kyritsakis, J. Paszkiewicz, S. Calatroni, W. Wuensch, and F. Djurabekova, Classification of vacuum arc breakdowns in a pulsed DC system, Phys. Rev. Accel. Beams 23, 023101 (2019).

[59] I. Profatilova, X. Stragier, S. Calatroni, A. Kandratsyeu, E. Rodriguez Castro, and W. Wuensch, Breakdown localisation in a pulsed DC electrode system, Nucl. Instrum. Methods Phys. Res., Sect. A 953, 163079 (2020).

[60] A. Descoeudres, T. Ramsvik, S. Calatroni, M. Taborelli, and W. Wuensch, DC breakdown conditioning and breakdown rate of metals and metallic alloys under ultrahigh vacuum, Phys. Rev. ST Accel. Beams 12, 032001 (2009).

[61] A. Lopez-Cazalilla, C. Serafim, J. Kimari, M. Ghaemi, A. Perez-Fontenla, S. Calatroni, A. Grudiev, W. Wuensch, and F. Djurabekova, Effect of surface orientation on blistering of copper under high fluence keV hydrogen ion irradiation, Acta Mater. 266, 119699 (2024).

[62] K. A. Esaklul, 13 - hydrogen damage, in *Trends in Oil and Gas Corrosion Research and Technologies*, edited by A. El-Sherik, Woodhead Publishing Series in Energy (Woodhead Publishing, Boston, 2017), pp. 315–340.

[63] H. Timko, M. Aicheler, P. Alknes, S. Calatroni, A. Oltedal, A. Toerklep, M. Taborelli, W. Wuensch, F. Djurabekova, and K. Nordlund, Energy dependence of processing and breakdown properties of Cu and Mo, Phys. Rev. ST Accel. Beams 14, 101003 (2011).

[64] G. Mesyats and V. Mesyats, The sequence of processes in the ecton cycle of a vacuum arc, J. Phys. Conf. Ser. 1115, 022020 (2018).